\documentclass[12pt,preprint]{aastex}

\shortauthors{Liu et al.}

\begin{document}

\title{Characteristics and Importance of ``ICME-in-Sheath" Phenomenon and Upper Limit for Geomagnetic Storm Activity} 

\author{Ying D. Liu\altaffilmark{1,2}, Chong Chen\altaffilmark{1,2}, and Xiaowei Zhao\altaffilmark{1,2}} 

\altaffiltext{1}{State Key Laboratory of Space Weather, National Space 
Science Center, Chinese Academy of Sciences, Beijing, China; liuxying@swl.ac.cn}

\altaffiltext{2}{University of Chinese Academy of Sciences, Beijing, China}

\begin{abstract}

As an important source for large geomagnetic storms, an ``ICME-in-sheath" is a completely shocked interplanetary coronal mass ejection (ICME) stuck in the sheath between a shock and host ejecta. Typical characteristics are identified from coordinated multi-sets of observations: (1) it is usually short in duration and lasts a few hours at 1 AU; (2) its solar wind parameters, in particular the magnetic field, seem to keep enhanced for a large range of distances; and (3) common ICME signatures are often lost. The host ejecta could be a single ICME or a complex ejecta, being fast enough to drive a shock. These results clarify previous misinterpretations of this phenomenon as a normal part of a sheath region. The ``ICME-in-sheath" phenomenon, together with a preconditioning effect, produced an extreme set of the magnetic field, speed and density near 1 AU in the 2012 July 23 case, all around their upper limits at the same time. This is probably the most extreme solar wind driving at 1 AU and enables us to estimate the plausible upper limit for geomagnetic storm activity. With an appropriate modification in the southward field, we suggest that a geomagnetic storm with a minimum $D_{\rm st}$ of about $-2000$ nT could occur in principle. The magnetopause would be compressed to about 3.3 Earth radii from the Earth's center, well inside the geosynchronous orbit. 

\end{abstract}


\section{Introduction}

The space weather community has long sought the ``worst-case scenario" that could possibly occur, as needed to benchmark space weather. In terms of geomagnetic storm intensity, the question boils down to what the most extreme value the $D_{\rm st}$ index can attain. Complete cancellation of the geomagnetic fields requires a $D_{\rm st}$ of about $-31,000$ nT \citep{parker67, riley18}, which is too extreme to be relevant. \citet{vasy11} suggests an upper limit of $-2500$ nT considering the limitation on the inflation of the magnetosphere by the Earth's dipole field at the equator. It is not clear, however, whether this limit could be achieved by the most intense solar wind driving possible at 1 AU.

Three solar wind parameters are of particular concern: the southward component of the interplanetary magnetic field $B_{\rm s}$, the solar wind speed $v$ and the density $n$ at 1 AU, in order of importance. These parameters produce the dawn-dusk electric field ($vB_{\rm s}$), which controls the rate of the solar wind energy coupling to the terrestrial magnetosphere \citep{dungey61}, and the solar wind dynamic pressure ($nv^2$), which governs the compression extent of the magnetosphere. Solar wind measurements of more than half a century indicate that a magnetic field of about 100 nT, a speed of about 2000 km s$^{-1}$ and a density of the order of 100 cm$^{-3}$ can be considered as their upper limits at 1 AU, respectively \citep[e.g.,][]{duston77, cliver90, skoug04, liu19}. How the three parameters achieve their respective upper limits at 1 AU simultaneously will provide important clues on the generation of the ``worst-case scenario". Coronal mass ejections (CMEs), large-scale expulsions of plasma and magnetic field from the corona, are the only candidate that may fulfill these upper limit conditions. Based on the observations of the 2012 July 23 complex CME, \citet{liu14a} suggest a ``perfect storm" mechanism for such simultaneous enhancements of the three parameters: preconditioning of the upstream medium by one or several earlier eruptions to prevent deceleration, plus in-transit interaction between later successive eruptions to preserve the magnetic field and density. \citet{liu15} extend the ``perfect storm" scenario to any combination of circumstances that can produce an event of unusual magnitude, not only the picture described above.  

A special case of the ``perfect storm" is a shock propagating inside a preceding CME or ICME (interplanetary counterpart of a CME). Here we term this phenomenon a ``shock-in-ICME" (SII). The shock enhances the pre-existing southward magnetic field within the ejecta, an idea for increased geo-effectiveness dating back several decades \citep{burlaga87, vandas97} with renewed interest in recent years \citep[e.g.,][]{liu14b, lugaz15a, lugaz15b, xu19, scolini20}. Note that shock compression can also amplify the solar wind speed and density at the same time, resulting in dynamic pressure enhancement in the sheath behind the shock. After the shock has crossed the preceding ejecta, the preceding ejecta will be stuck inside the sheath between the shock and its driver (which we call the ``host ejecta"). This is what we define an ``ICME-in-sheath" (IIS) here\footnote{\citet{lugaz05} describe various interaction phases between two CMEs using numerical simulations. Our IIS phenomenon corresponds to their leading ejecta structure that has been completely shocked after the end of the shock-ejecta interaction phase. The sheath of the preceding ejecta also becomes part of the sheath of the host ejecta, although the shock may be still propagating in the sheath of the leading ejecta (see the 2001 November case below). Note that the IIS phenomenon is usually neglected in reality because it often appears as turbulent as a normal sheath from in situ measurements.}. The difference between a SII and an IIS is non-trivial in terms of space weather. First, a SII leads to only shock compression of the preceding ejecta, while for an IIS the preceding ejecta is further squeezed by the host ejecta on top of shock compression. Second, to have enhanced geo-effectiveness a SII has to ``happen to" occur at 1 AU, but enhancements in the field, speed and density are locked inside an IIS for a wide range of distances (see Section 2). Another reason we stress the IIS phenomenon is that it is often misinterpreted as a normal part of the sheath region, so it is necessary to make clarifications. Our efforts here echo the importance of sheaths for large geomagnetic storms \citep[e.g.,][]{tsurutani92, kilpua17}. The nature of an IIS, however, is a totally shocked ICME entrained in the sheath between the shock and host ejecta, so its geo-effectiveness would be even higher than a normal sheath.    

In addition to providing a clear definition of the IIS phenomenon, a focus of this Letter is to highlight IIS characteristics and importance for space weather. We illustrate how the IIS phenomenon and a preconditioning effect result in probably the most extreme solar wind driving at 1 AU, which is then used to estimate the plausible upper limit for geomagnetic storm activity. Our results also clarify previous misinterpretations about this important phenomenon, which may lead to new physics concerning the propagation of CMEs.  

\section{Observations and Results} 

Table~1 summarizes the parameters of six IIS examples. The purpose is not to provide a comprehensive survey of IISes, but to illustrate their typical characteristics that can facilitate future identification and to highlight their importance for space weather. Note that an unambiguous identification of the IIS phenomenon requires remote-sensing observations (e.g., wide-angle imaging or a long-duration type II radio burst) in connection with in situ measurements. These cases are carefully picked so that we can follow how the patterns of interacting CMEs evolve with time and how the interaction features connect with in situ signatures. A first impression from Table~1 is that an IIS is usually short in duration, lasting a few hours at 1 AU, but has a much larger magnetic field than the host ejecta. Three examples are described below in detail covering some complexity and diversity of the IIS phenomenon. As for the remaining cases, we refer readers to \citet{liu12} for the 2010 August case \citep[with additional information from][]{harrison12, mostl12, juan12, webb13}, \citet{liu13} for the 2012 January case, and \citet{liu19} for the 2017 July case. Although not described here, they can be readily understood when put in the context of the present work.

\subsection{The 2001 November Case}

The 2001 November case is selected here, because a long-duration type II burst, in combination with coronagraph and multi-point in situ observations, provides key features for determining the propagation and interaction properties of three successive CMEs over large distances. Readers are directed to \citet{liu17} for a detailed analysis, and additional information can be found in \citet{rodriguez08} and \citet{reisenfeld03}.

Figure~1 shows the in situ measurements at the Earth. The first ICME appears in the sheath (IIS) of the host ejecta, which is a complex ejecta \citep{burlaga02} formed from the merging of the later two CMEs. The second shock in Figure~1, after its survival from the CME merging, has passed through the IIS at 1 AU and is about to merge with the shock driven by the IIS. Inside the IIS, the magnetic field, speed and density (as well as the temperature) are greatly enhanced. This is due to shock compression as it was propagating through the IIS, plus further compression by the host ejecta from behind. The southward magnetic field is as high as 50.5 nT within the IIS. Although only for a brief time period, it caused an intense geomagnetic storm (minimum $D_{\rm st}$ of $-221$ nT) together with the high speed. The IIS does not have typical ICME signatures \citep[for the typical signatures at 1 AU see][]{zurbuchen06}, which are presumably lost in the interaction process. This is why it is often misinterpreted as a normal part of a sheath region. While showing an indication of rotation, the field components are turbulent. There is a reduction in the proton $\beta$ within the IIS compared with the surrounding medium, but on average it is much higher than that in the host ejecta. Ulysses measurements at 2.34 AU provide further evidence for the IIS, whose field enhancement is still maintained there \citep[see Figure~4 in][]{liu17}. These coordinated measurements reveal an important conclusion concerning the evolution of an IIS that, once stuck in the sheath between the shock and host ejecta, its field, speed and density would keep enhanced for a long time. 
 
\subsection{The 2011 February Case}

The 2011 February case is chosen, as wide-angle imaging observations from STEREO show how a pileup of successive CMEs lead to an IIS. During the time period of February 13-15, the Sun exhibited substantial activity including the first X-class flare of solar cycle 24 from NOAA AR 11158 (S20$^{\circ}$W10$^{\circ}$) on February 15. With the two STEREO spacecraft separated by about 180$^{\circ}$, a 360$^{\circ}$ view of the Sun was achieved for the first time. The same active region produced a series of CMEs on February 13-15 moving roughly along the Sun-Earth line, as can be seen from Figure~2. At least 3 CMEs merge into a single track. As will be shown later, this is recognized as a shock from the connection between imaging and in situ observations. We focus on this track and use a triangulation technique to convert its elongation angles to kinematics \citep{liu10}. An average propagation angle of about 3.5$^{\circ}$ east of the Sun-Earth line is derived from the analysis. A linear fit of the resulting distances gives a speed of about 600 km s$^{-1}$ and a predicted arrival time of about 23:15 UT on February 17 at the Earth.

The in situ measurements near the Earth are shown in Figure~3. A forward shock passed Wind around 00:49 UT on February 18 with a speed of about 530 km s$^{-1}$, which agrees with the predictions from imaging observations. This agreement confirms that the track after the CME merging in Figure~2 represents the shock. The IIS, again, is short and characterized by an enhanced magnetic field, speed and density. A reasonable explanation is shock compression during its passage through the preceding ejecta, with additional push by the host ejecta from behind. The temperature inside the IIS, although lower than the expected one, is higher than the ambient temperature ahead of the shock, indicative of heating during the IIS compression. The IIS field components are still turbulent to some extent. In this case the proton $\beta$ is much reduced within the IIS, which suggests that the IIS may have a different nature compared with a normal sheath. The host ejecta is also a complex ejecta, as can be seen from the multiple depressions in the temperature. No intense geomagnetic storm occurred despite an obvious sudden commencement in the $D_{\rm st}$ index. This is probably because the field lacks a persistent, strong southward component in the IIS and host ejecta. More discussions on the CME-CME interaction can be found in \citet{maricic14} and \citet{temmer14}. 

\subsection{The 2012 July Case}

The 2012 July case is an extreme one, which combines a preconditioning effect leading to an unusually high speed and in-transit interaction between two consecutive eruptions resulting in an IIS \citep{liu14a}. Figure~4 shows the in situ measurements at STEREO A near 1 AU. We observe IIS characteristics similar to those of the cases discussed above, but in this case the host ejecta is a single ICME. The proton $\beta$, which is reduced within the IIS, can only be estimated based on the expected temperature, as the measurements of the proton temperature are largely missing across the shock and the IIS and host ejecta periods. Inside the IIS the magnetic field is extremely enhanced, with a peak value around 109.2 nT and an average of about 90.3 nT. The speed is 2246 km s$^{-1}$ right behind the shock and has a mean value close to 1700 km s$^{-1}$ in the IIS. Electron measurements suggest that the plasma density in the IIS may have been as high as $\sim$150 cm$^{-3}$. All these are record values at 1 AU. The magnetic field within the host ejecta also remains very high at 1 AU, as its expansion is inhibited by the interaction between the two ICMEs. We refer readers to \citet{liu14a} for more discussions and other studies \citep[e.g.,][]{russell13, baker13, riley16} for supplementary information.      

Clearly, the IIS phenomenon together with the preconditioning effect gives rise to an extreme set of the solar wind magnetic field, speed and density at 1 AU, all around their upper limits at the same time (we have indicated in Section 1 the respective upper limits of the three parameters based the historical data at 1 AU). This is probably the strongest solar wind forcing at 1 AU. We use this set of solar wind parameters (with an appropriate modification) to estimate the upper limit for geomagnetic storm activity. The only modification made here is the southward field within the two ejecta and their interface, for which we take the negative value of the field magnitude. This essentially re-orients the axes of the CME flux ropes, with assumed dominant axial fields, to be purely southward. The magnetic field at the axis of any flux rope is entirely axial (i.e., no azimuthal components) and usually the largest over its cross section. Therefore, southward fields approximating to the field magnitudes can be obtained at least near the axis for a completely southward flux rope. If the axial fields predominate over the azimuthal components throughout the flux rope, we will have a case similar to the assumed here. 

The dawn-dusk electric field would be as high as $\sim$180 mV m$^{-1}$. We evaluate the $D_{\rm st}$ index using two empirical formulas \citep{burton75, om00} based on the modified southward field. Note that these models have been calibrated on more modest geomagnetic storms. Application to the present extreme case should be taken with caution. Below we compare different approaches in order to get a reliable estimate. Our experience indicates that the \citet{burton75} model tends to overestimate, while the \citet{om00} scheme tends to underestimate, the $D_{\rm st}$ index, so we average their results. The resulting $D_{\rm st}$ profile has a minimum value of about $-2000$ nT. On the other hand, an empirical relation of $D_{\rm st}=-0.01vB_{\rm s}-25$ nT \citep{gopal08} would give about the same value for $v\sim2000$ km s$^{-1}$ and $B_{\rm s}\sim100$ nT, if its extrapolation to an extreme case is valid. Our result seems smaller than the estimate of \citet{vasy11}, i.e., $-2500$ nT, but not significantly so.  

Figure~4 also indicates that the maximum solar wind dynamic pressure in the IIS (and the surrounding sheath as well) may have exceeded 700 nPa, about 450 times the quiet-time level ($\sim$1.6 nPa). An empirical model \citep{shue98} suggests that the dynamic pressure together with the assumed southward field would compress the subsolar magnetopause, typically located at $\sim$10 Earth radii, to about 3.3 Earth radii. This is well inside the geosynchronous orbit. 

\section{Conclusions and Discussion}

We have highlighted the IIS phenomenon, defined as a completely shocked ICME stuck in the sheath between a shock and host ejecta, and its importance for space weather. An IIS is anticipated to occur frequently especially near solar maximum. Through coordinated multi-sets of observations, examples have been carefully identified covering complexity and diversity of this phenomenon. Typical characteristics are found. First, an IIS is usually short and lasts a few hours at 1 AU. Second, enhanced solar wind parameters, in particular the magnetic field, are observed inside an IIS, and the enhancements may persist for a long time. This is attributed to compression by shock passage as well as the host ejecta. Third, common ICME signatures are often lost. A structure resembling a small magnetic cloud, however, may be seen occasionally \citep[e.g., the 2017 July case in][]{liu19}. As in \citet{liu12}, we suggest that a depressed proton $\beta$ may be useful for the IIS identification, as the compression may not change the plasma $\beta$ much. The host ejecta could be a single ICME or a complex ejecta formed from merging of multiple CMEs. 

Given the locked, simultaneous enhancements in the solar wind parameters, an IIS is an important source for large geomagnetic storms. On average its geo-effectiveness is expected to be higher than a normal sheath, ICME and even SII. In some of the cases we observe a large geomagnetic storm or infer one if the case had been Earth directed. However, for some other cases an intense geomagnetic storm did not occur (e.g., the 2011 February case). This really depends on whether a persistent, strong southward field component exists, although the magnetic field magnitude is generally high inside an IIS. Due to the increased internal pressure, the enhancement in the IIS field may relax after the end of the shock-ejecta interaction phase in the light of numerical simulations \citep{lugaz05, xiong07, scolini20}. This should be a slow process, as the IIS expansion is confined by the host ejecta from behind. The 2001 November case, for which multi-point in situ measurements are available, indicates that the field enhancement can last long in an IIS.

A plausible estimate of the most extreme solar wind driving at 1 AU is obtained, based on the 2012 July case with an appropriate modification in the southward field. The IIS phenomenon in combination with a preconditioning effect, in this case, leads to an extreme set of the magnetic field, speed and density that are all around their upper limits at the same time. The magnetopause would be compressed to about 3.3 Earth radii from the Earth's center, well below the geosynchronous orbit. Many satellites, such as those for communication, navigation, positioning and weather, would be under direct impacts of the solar wind and energetic particles without protection of the magnetosphere. In principle, a geomagnetic storm with a minimum $D_{\rm st}$ of about $-2000$ nT could occur given the observed field and speed. These probably set the upper limit for geomagnetic storm activity. It is not that the magnetosphere saturates at $D_{\rm st}\sim-2000$ nT, but rather the solar wind is not capable of producing a higher forcing at 1 AU. For comparison, the 1859 Carrington storm, a well-known example of extreme space weather before the space era, reached about $-850$ nT \citep{siscoe06}, whereas the 1989 March storm, the most severe of the space age, arrived at only $-589$ nT.

\acknowledgments The research was supported by NSFC under grant 41774179, Beijing Science and Technology Plan Project (Z191100004319003), and the Specialized Research Fund for State Key Laboratories of China. We acknowledge the use of data from STEREO and Wind and the $D_{\rm st}$ index from WDC in Kyoto. Readers are directed to \citet{kaiser08} for the STEREO mission, \citet{ogilvie95} for Wind/SWE, and \citet{lepping95} for Wind/MFI.

\clearpage

\begin{deluxetable}{lccccc}
\tabletypesize{\small}
\tablecaption{Estimated Parameters from In Situ Measurements near 1 AU}
\tablewidth{0pt}
\tablehead{
\colhead{Date} & \colhead{Shock\tablenotemark{a}} & \colhead{IIS\tablenotemark{a}} 
& \colhead{Host ejecta\tablenotemark{a}} & \colhead{$B_{\rm I}$\tablenotemark{b}} & 
\colhead{$B_{\rm H}$\tablenotemark{b}} \\
& (UT) & (UT) & (UT)  & (nT) & (nT)}
\startdata
2001 Nov &Nov 24 05:53 &Nov 24 07:41 + 4.1 hr &Nov 24 16:34 + 26.9 hr &44.2 &14.3 \\
2010 Aug &Aug 3 17:02   &Aug 4 03:07 + 4.1 hr   &Aug 4 09:50 + 14.9 hr   &15.3 &8.7   \\
2011 Feb &Feb 18 00:49 &Feb 18 04:19 + 4.6 hr  &Feb 18 19:55 + 36.5 hr &27.2 &9.7   \\
2012 Jan &Jan 22 05:33 &Jan 22 11:31 + 6.2 hr   &Jan 22 23:38 + 39.0 hr  &20.7 &7.3  \\
2012 Jul  &Jul 23 20:55  &Jul 23 22:58 + 2.4 hr    &Jul 24 03:22 + 19.8 hr   &90.3  &28.4 \\
2017 Jul  &Jul 24 17:54  &Jul 24 22:48 + 5.1 hr    &Jul 25 07:15 + 51.7 hr   &45.9  &4.8   \\
\enddata
\tablenotetext{a}{Shock arrival time, and the leading edge plus duration of the IIS/host ejecta, respectively.}
\tablenotetext{b}{Average magnetic field strength inside the IIS/host ejecta, respectively.}
\end{deluxetable}

\clearpage

\begin{figure}
\epsscale{0.7} \plotone{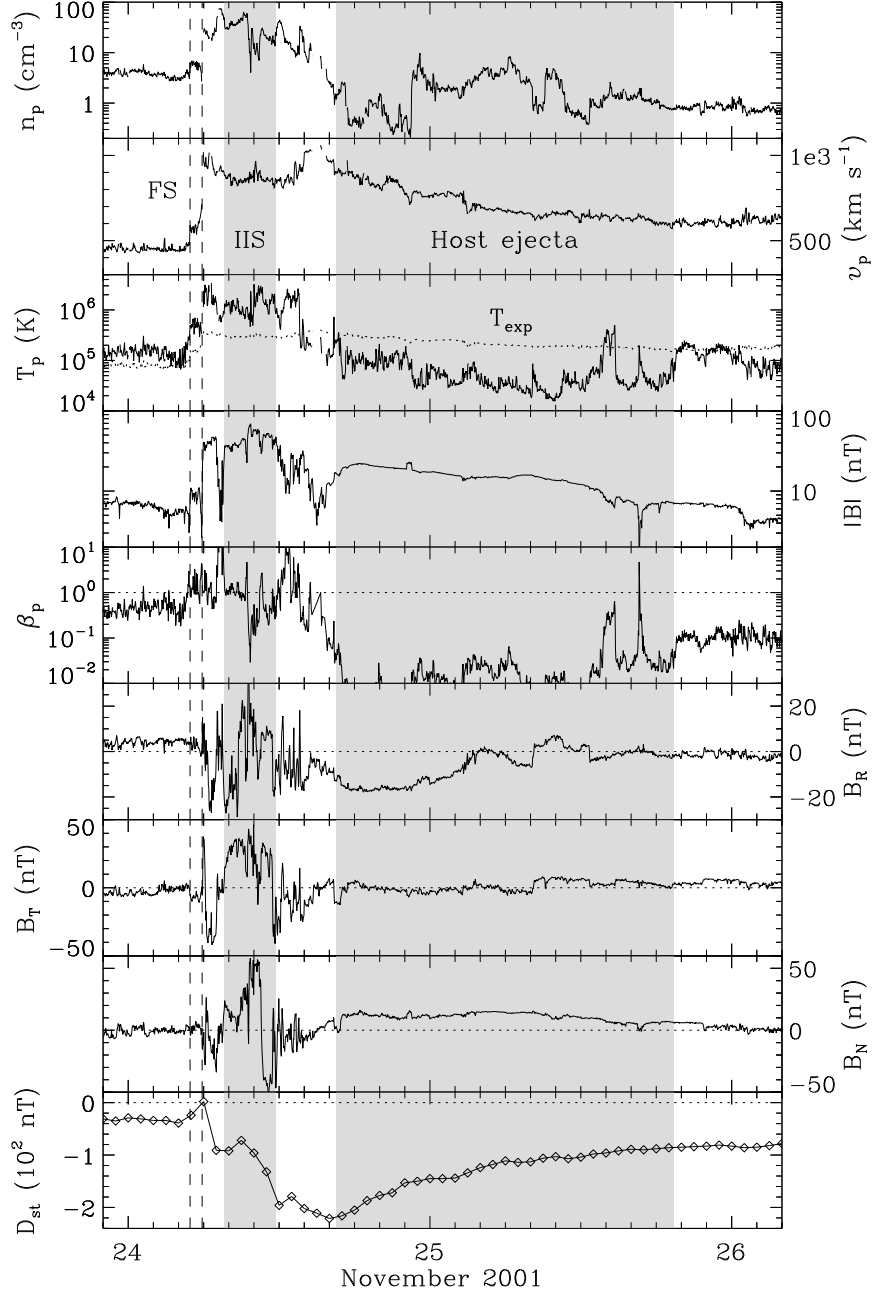} 
\caption{Solar wind measurements at Wind and $D_{\rm st}$ index for the 2001 November case. From top to bottom, the panels show the proton density, bulk speed, proton temperature, magnetic field strength, proton $\beta$, magnetic field components, and $D_{\rm st}$ index, respectively. The shaded regions represent the intervals of the IIS and host ejecta, and the vertical dashed lines mark their respective forward shocks (FS) that are about to merge. The dotted curve in the third panel denotes the expected proton temperature calculated from the observed speed \citep{lopez87}.}
\end{figure}

\clearpage

\begin{figure}
\epsscale{0.7} \plotone{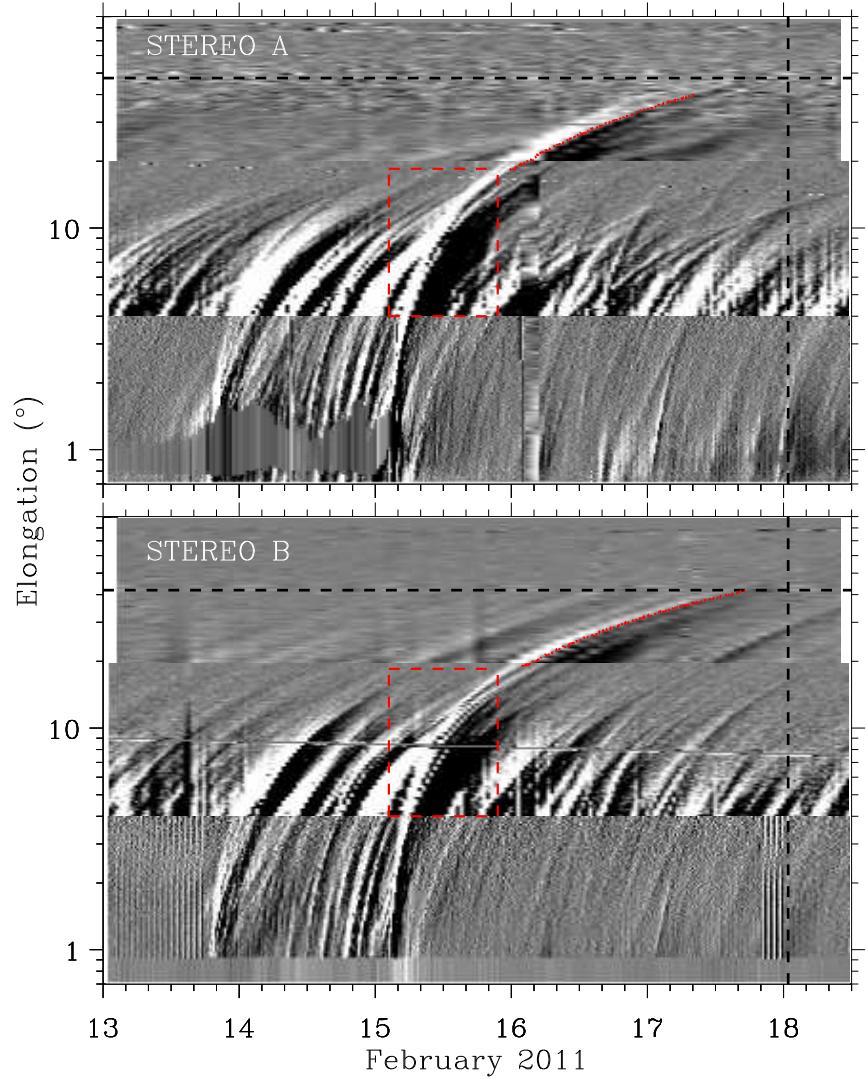} 
\caption{Time-elongation maps constructed from running-difference images of COR2, HI1 and HI2 along the ecliptic plane for STEREO A (upper) and B (lower). The rectangular box marks the merging of at least 3 CMEs into a track (red dotted curve), which later likely represents a shock wave. The vertical dashed line indicates the observed arrival time of the shock at the Earth. The horizontal dashed line denotes the elongation angle of the Earth.}
\end{figure}

\clearpage

\begin{figure}
\epsscale{0.7} \plotone{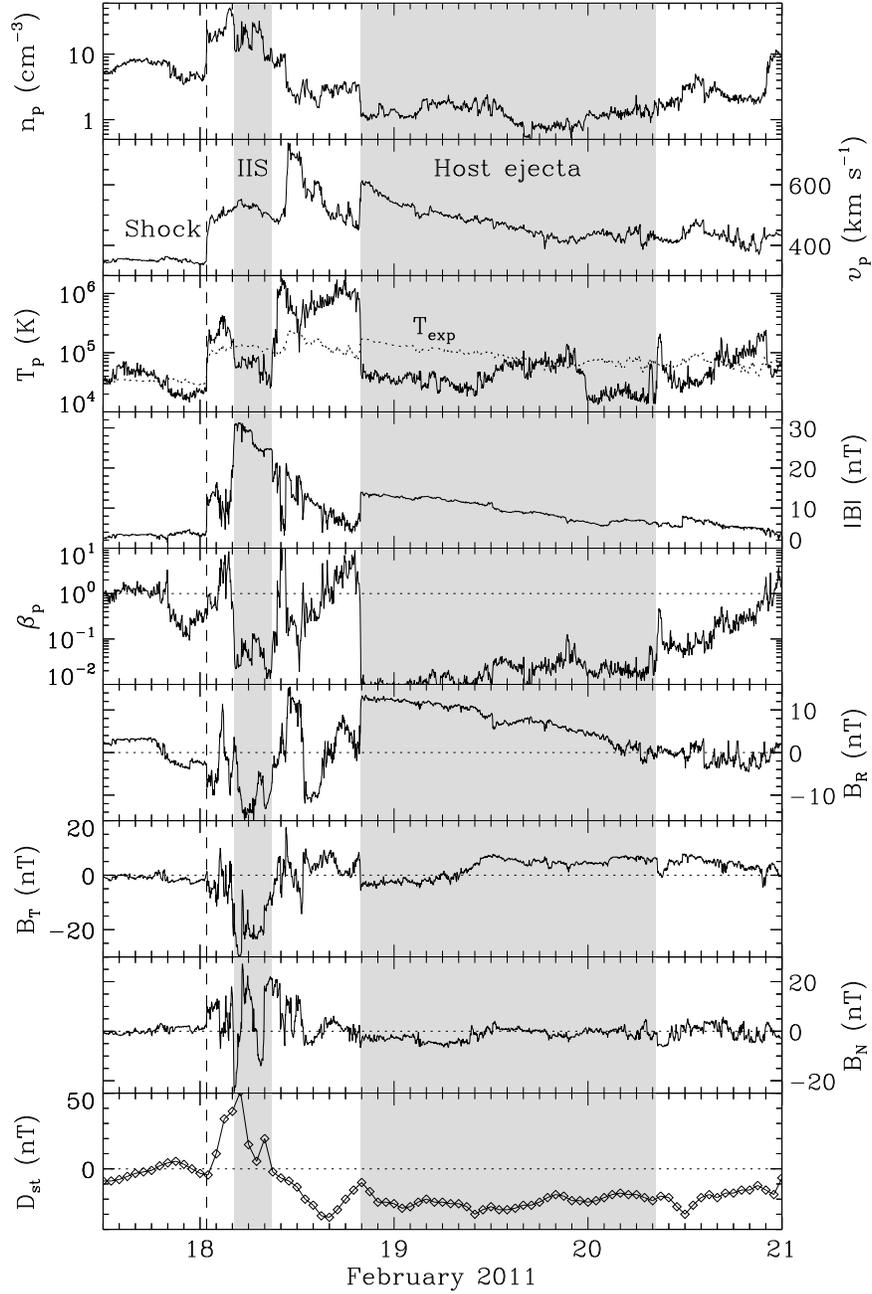} 
\caption{Solar wind measurements at Wind and $D_{\rm st}$ index for the 2011 February case. Similar to Figure~1.}
\end{figure}

\clearpage

\begin{figure}
\epsscale{0.7} \plotone{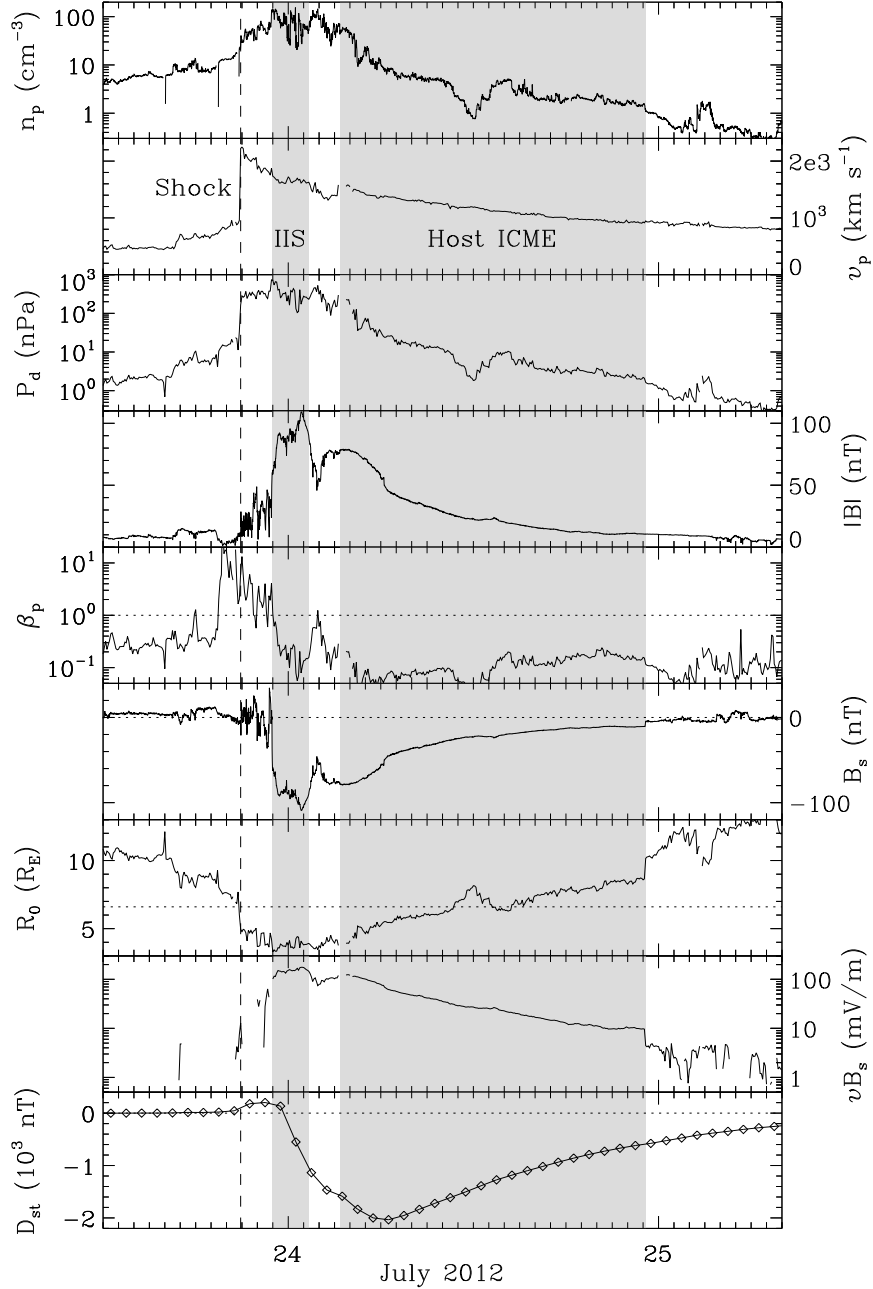} 
\caption{Solar wind parameters associated with the 2012 July extreme case used to estimate the upper limit of geomagnetic storm activity. From top to bottom, the panels show the number density estimated from electron measurements \citep{liu14a}, bulk speed, dynamic pressure, magnetic field strength, proton $\beta$ estimated from the expected temperature, assumed southward field component (by inverting the field magnitude inside the ejecta and their interface), calculated subsolar distance of the magnetopause from the Earth's center, dawn-dusk electric field, and simulated $D_{\rm st}$ index, respectively. The horizontal dotted line in the seventh panel indicates the geosynchronous orbit altitude (6.6 Earth radii).}
\end{figure}

\end{document}